\documentclass[prd,aps,floats,epsfig,twocolumn]{revtex4}
\usepackage{graphicx}
\newcommand{\beq}{\begin{equation}}
\newcommand{\eeq}{\end{equation}}
\newcommand{\bea}{\begin{eqnarray}}
\newcommand{\eea}{\end{eqnarray}}

\newcommand{\ApJ}{{\it Astrophys. J.\,}}

\newcommand{\PR}{{\it Phys. Rev.\,}}

\newcommand{\etal}{{\it et al.}}
\begin{document}

\title{Recombining WMAP: \\ constraints on ionizing and resonance
radiation at recombination}
\author{Rachel Bean$^\sharp$, Alessandro Melchiorri$^{\flat,*}$ and Joe Silk
$^\flat$}
\affiliation{$^\sharp$ Dept. of Astrophysical Sciences, Princeton University, 
Princeton, NJ 08544, USA.\\$^\flat$ Astrophysics, Denys Wilkinson Building, University of Oxford, Keble
road, OX1 3RH, Oxford, UK\\$^*$ Universita' di Roma ``La Sapienza, Ple Aldo Moro 2, 00185, Rome, Italy}

\begin{abstract}

We place new constraints on sources of ionizing and 
resonance radiation at the epoch of the recombination process using the recent CMB temperature and polarization spectra 
coming from WMAP.  
We find that non-standard recombination scenarios are still consistent with the current data.
In light of this we study the impact that  such models can 
have on the determination of several cosmological parameters.
In particular, the constraints on curvature and baryon density appear to be
weakly affected by a modified recombination scheme. However, it may affect the
current WMAP constraints on inflationary parameters like the spectral
index, $n_{s}$, and its running. Physically motivated models, like those based on 
primordial black hole or super heavy dark matter decay, are able
to provide a good fit to the current data.
Future observations in both temperature and polarization
will be needed to more stringently test these models.

\end{abstract}
\maketitle

\section{Introduction}
The recent measurements of the Cosmic Microwave Background (CMB) flux provided 
by the Wilkinson Microwave Anisotropy Probe (WMAP) 
mission \cite{Bennett03} have truly marked the beginning of the era of 
precision cosmology. In particular, the position and amplitude of the
 detected oscillations in the angular power spectrum of the CMB 
are in spectacular agreement
with the expectations of the standard model of structure formation,
based on primordial adiabatic and nearly scale invariant perturbations
\cite{page}.  Assuming this model of structure formation, an 
{\it indirect} but accurate measurement of several 
cosmological parameters has been reported \cite{spergel} in agreement with those previously indicated  (see e.g. \cite{tegmark02} and \cite{cjo}). 

However, beside a full confirmation of the previous scenario 
(with a sensible reduction of the error bars!)
the WMAP data set is also hinting towards a modification of the standard 
picture in several aspects \cite{spergel}.
In particular, the low CMB quadrupole \cite{contaldi,efstathiou}
possible scale dependence of the spectral index \cite{peiris,trodden}, 
high optical depth (see e.g. \cite{kogut,ciardi,chiu}) possible 
deviations from flatness \cite{uzan} and dark energy
(\cite{melchiorri}) have already produced a wide interest. 
 Further, the relatively high $\chi^2$ per degree of freedom of the 
fiducial model suggests systematics and/or new physics could be 
considered. 

It is therefore timely, with the increased precision of the 
CMB dataset to further test the standard scenario and to investigate 
possible deviations.

Here we investigate possible deviations in the mechanism to which CMB 
anisotropies are higly dependent: the process of recombination. 

The recombination process can be modified in several ways. For example, one could use a model independent, phenomenological approach such as in \cite{Hannestad01} where models are specified by the position and width of the recombination surface in redshift space. Here
we will instead focus on theoretically motivate mechanisms based on extra sources of ionizing and 
resonance radiation at recombination (see e.g. \cite{Seager00}).
While the method we adopt will be general enough to cover
most of the models of this kind, we remind the
reader that there are several other ways in which to modify 
recombination, like, for instance, by having a time-varying 
fine-structure constant (\cite{alpha}).
In our analysis, we will not cover the constraints on these models from WMAP as they have been recently investigated in (\cite{carlos}). 

Following the seminal papers \cite{Peebles68,Zeldovich68} detailing the 
recombination process, further refinements to the standard scenario 
were developed \cite{Seager99} allowing predictions at the accuracy level 
found in data from the WMAP satellite and the  future Planck satellite 
\cite{Hu95,Seljak03}. With this level of accuracy it becomes conceivable 
that deviations from standard recombination maybe be detectable 
\cite{Seager00,Naselsky02,Dorosh02}.

The letter proceeds as follows: in section \ref{sec2} we review models 
which can produce deviations from the standard recombination scenario. 
In \ref{sec3} we describe how these deviations might affect the 
CMB temperature and polarization power spectra and conduct a likelihood 
analysis using the recent CMB data from WMAP. 
In particular, we will study the impact that a modified recombination 
scheme can have on several cosmological parameters. 
In \ref{sec4} we draw together the implications of the analysis, 
placing constraints on current theories of recombination.

\section{Beyond standard recombination}\label{sec2}

In the standard recombination model  \cite{Peebles68,Zeldovich68} the 
net recombination rate is given by 
\bea 
-{dx_{e}\over dt}\left.\right|_{std}=C\left[a_c n x_{e}^{2}-b_c 
(1-x_{e})\exp{\left(-{\Delta B\over k_{B}T}\right)}\right]
\label{eq1}
\eea
where $x_{e}$ is the ionization fraction,  $a_{c}$ and $b_{c}$ are 
the effective recombination and photo-ionization rates for principle 
quantum numbers $\ge 2$, and $\Delta B$ is the difference in binding energy between the $1^{st}$ and $2^{nd}$ energy levels. In addition to the single Ly-$\alpha$ transition with wavelength $\lambda_{\alpha}$, there is also 2 photon decay 
from the meta-stable 2s level, with decay rate $\Lambda_{1s2s}$. 
The contribution of this process is reflected in the parameter $C$,
\bea C={1+K\Lambda_{1s2s}n_{1s}\over1+K(\Lambda_{1s2s}+b_{c})n_{1s}}
\label{eqC},  \ \ \ \ K={\lambda_\alpha^{3} \over8\pi H(z)}
\eea

The standard hydrogen recombination 
scenario can be simply extended in two ways with the addition of Ly-$\alpha$ 
and ionizing photons  \cite{Seager00}. 

The extra contributions can be related to the baryon number density through 
the efficiency functions $\varepsilon_{\alpha}$ and $\varepsilon_{i}$, 
respectively,
\bea
{dn_{\alpha}\over dt}&=&\varepsilon_{\alpha}(z)H(z)n, \ \ \ 
{dn_{i}\over dt}=\varepsilon_{i}(z)H(z)n 
\label{eq0}
\eea
where $H(z)$ and $n$ are the Hubble expansion parameter and mean baryon 
density (neutral H and protons) respectively. We leave photons produced by two body interactions, for example of annihilation of WIMPS, with $dn_{i}/dt=\sigma_{WIMP}n_{WIMP}^{2}$ for future investigation.

The addition of extra Ly-$\alpha$ and ionization photons  in equation 
(\ref{eq0}) adjusts the recombination rate from the standard model 
in equation (\ref{eq1}) 
\bea -{dx_{e}\over dt}&=&-{dx_{e}\over dt}\left.
\right|_{std}-C\varepsilon_{i} H-(1-C)\varepsilon_{\alpha}H . 
\ \ \ \ \ \ \ \label{eq2}
\eea

There are a range of physical mechanisms which could generate additional 
photons, we review them briefly here, see \cite{Dorosh02} and 
references therein for a fuller discussion and derivations.

\begin{figure}[t]
\begin{center}
\includegraphics[width=3in]{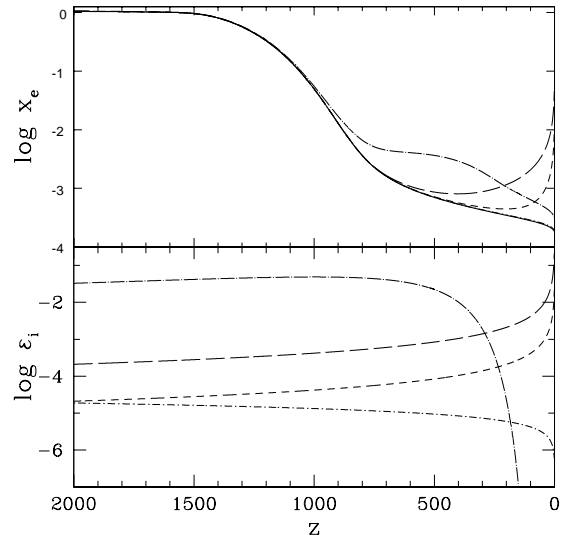}
\caption{Evolution of $\varepsilon_{i}$, characterizing the number density of extra ionizing photons added during the recombination epoch (bottom panel) and the evolution of the associated ionization fraction $x_{e}$ (top panel) for the WMAP best fit, fiducial case (full line, $\varepsilon_{i}=0$) and
for sample models, described in 
the text: PBH decay with coefficients as in \cite{Dorosh02} (dot-long dash), 
topological defect particle release (dot-short dash),  short lived SHDM decay 
with $\Theta=10^{5}$ (short dash), and $\Theta=10^{6}$ (long dash) 
corresponding to $z(t_{x})\sim 4 $ and 5 respectively. The ionization fraction for the topological defect model lies almost on top of the fiducial case. The parameter measuring additional resonance photons, $\varepsilon_{\alpha}$,  is of the same order as$\varepsilon_{i}$ in each of these models.}\label{fig1}
\end{center}
\end{figure}

In general $\varepsilon_{\alpha}$ and $\varepsilon_{i}$ are functions of 
redshift, dependent upon the particle decay model that is being employed, 
and the size of the decay lifetime, $t_{x}$ in comparison to the lifetime of 
the universe $t_{0}$. 

Primordial black hole (PBH) decay \cite{Naselsky87} at the moment of 
recombination can be described by a model of the form
\bea
\varepsilon_{j}= c_{j}\zeta^{-{3\over 2}}\exp{-\zeta^{-{3\over 2}}}, 
\ \ \ \zeta={1+z\over 1+z_{dec}}
\eea
where $j$ can be $\alpha$ or $i$ and $z_{dec}$ is the redshift of photon 
decoupling. The authors of \cite{Dorosh02} use $c_{\alpha}\approx 0.3$ 
and $c_{i}\approx 0.13$ however these values are dependent on assumptions 
about the PBH mass and number density distributions, which in turn rely on 
the nature of the inflationary spectrum \cite{Liddle98} and the possibility 
of accretion \cite{Bean02}.

\begin{figure}[t]
\begin{center}
\includegraphics[width=3in]{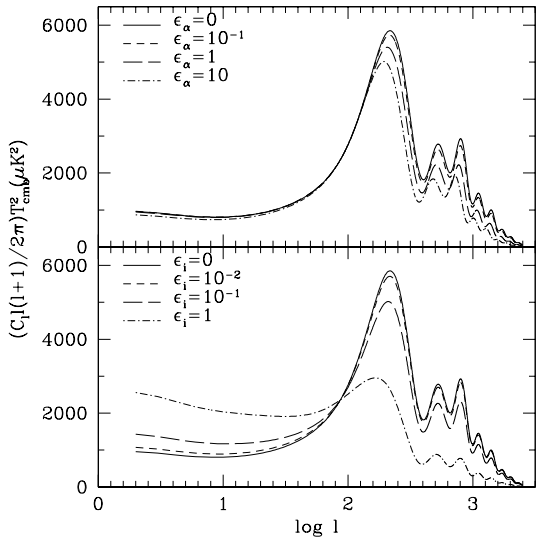}
\caption{CMB TT spectra for various values of $\varepsilon_{\alpha}$ 
(top) and $\varepsilon_{i}$ (bottom) showing the increasing suppression and shift of the peaks as one increases the number of extra resonance and ionizing photons, respectively, in comparison to the WMAP best fit fiducial model with $\varepsilon_{i}=\varepsilon_{\alpha}=0$ (full line). Spectra are normalized to $C_{87}$.}  \label{fig2}
\end{center}
\end{figure}

Electromagnetic cascades from general particle decay can been described by 
the parameterization \cite{Dorosh02} 
\bea
\varepsilon_{\alpha,i}\sim{10^{-8}\over\Omega_{b}^{0}h^{2}}\Theta[z(t_{x})] 
(1+z)^{r}, 
\eea
with $r=1/2$ for particle release from topological defects, $r=-1$ for decay 
of super heavy dark matter (SHDM), and where the normalization is the 
EGRET energy density \cite{Berton01}, although one could imagine using a code such as DARKSUSY \cite{DARKSUSY} to get a more precise normalization value. The proportionality constant 
$\Theta[z(t_{x})]$ reflects an alteration required for decay time dependence 
in the normalization of SHDM scenarios. For short lived SHDM with 
$t_{x}< t_{0}$, 
$\Theta[z(t_{x})]\sim\exp{\left[\left(1+z[t_{x}]\right)^{3/2}\right]}$ 
and equals 1 for all other cases. Short lived SHDM can therefore have 
substantially higher values for $\epsilon_{i,\alpha}$. 

Note that in both particle and BH decay the models predict  
$\epsilon_{i}\sim\epsilon_{\alpha}$. We demonstrate the size and 
evolution of $\epsilon_{i}$ for scenarios described above in Fig. \ref{fig1}. 

The combined effect on reionization from mechanisms such as those described 
above would obviously require a sum over contributions from all sources. 
Our aim is to find an upper limit on the overall contribution, independent 
of source, and as such we adopt a simple parameterization using constant, 
effective values for $\varepsilon_{i}$ and $\varepsilon_{\alpha}$.

\begin{figure}[t]
\begin{center}
\includegraphics[width=3in]{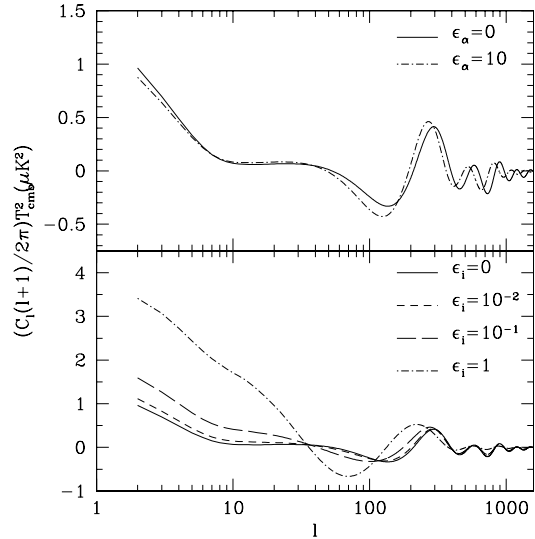}
\caption{CMB TE spectra for various values of $\varepsilon_{\alpha}$ (top) 
and $\varepsilon_{i}$ (bottom) against a fiducial case with  $\varepsilon_{\alpha}= \varepsilon_{i}=0$ (full line). In each case a reionization redshift of 
$z_{ri}=7$ is used for comparison.  }
\label{fig3}
\end{center}
\end{figure}

In Figs.~\ref{fig2} and \ref{fig3} we plot the temperature (TT) and 
temperature-polarization cross correlation (TE) power spectra for several 
values of $\varepsilon_{\alpha}$ and $\varepsilon_{i}$. Qualitatively, 
increasing $\varepsilon_{\alpha}$ broadens the epoch of recombination and 
lowers the redshift when the optical depth drops through unity.  
Delaying recombination increases the angular diameter distance at last 
scattering and therefore shifts the first peak to lower $l$. 
The delay also increases the optical depth, suppressing the height of the 
peaks in comparison to the low $l$ plateau and decreasing the ratio of 
the second to first peaks. 

Increasing $\varepsilon_{i}$ introduces a plateau in ionization fraction, 
preventing a trend towards full recombination. 
Boosting the number density of ionizing photons, increases the optical 
depth significantly more than a similar increase in Ly-$\alpha$ photons, 
resulting in a pronounced suppression on the first peak height. 
Subsequently we would expect much tighter constraints on $\varepsilon_{i}$ 
from the TT and TE spectra.

It is interesting to note that introducing a large extra ionizing component can give a large 
cross-correlation on large scales, similar to that generated by early 
reionization. However it is unable to account for both the TT and TE 
observations simultaneously because of the peak suppression in the TT spectrum.

\section{Likelihood analysis}
\label{sec3}
Our analysis method is based on the computation of a likelihood 
distribution over a grid of precomputed theoretical models. 

We restrict our analysis to a flat, adiabatic, $\Lambda$-CDM model 
template computed with a modified version of CMBFAST \cite{CMBFAST}, 
sampling the parameters as follows: 
$\Omega_{cdm}h^2\equiv \omega_{cdm}= 0.05,...0.25$, in steps of $0.01$; 
$\Omega_{b}h^2\equiv\omega_{b} = 0.009, ...,0.030$, in steps of $0.001$ and 
$h=0.55, ..., 0.85$, in steps of $0.05$. 
The value of the cosmological constant $\Lambda$ is determined by the 
flatness condition.
Our choice of the above parameters is motivated by Big Bang Nucleosynthesis 
bounds on $\omega_b$ (both from $D$ \cite{burles} and $^4He+^7Li$ 
\cite{cyburt}), from supernovae \cite{super1} and galaxy clustering 
observations (see e.g. \cite{thx,hansen}). 
From the grid above we only consider models with age of the universe 
$t_0>11$ Gyrs.
We vary the spectral index of the primordial density perturbations within 
the range $n_s=0.8,..,1.2$, we allow for a possible (instantaneous) 
reionization of the intergalactic medium by varying the reionization redshift 
$5 < z_{ri}< 25$  and we allow a free re-scaling of the fluctuation 
amplitude by a pre-factor of the order of $C_{87}$, in units of 
$C_{87}^{NORM}= 1.9 \mu K^2$ 
Finally, we let $\varepsilon_{\alpha}$ and $\varepsilon_{i}$ vary as follows: 
$10^{-4}<\varepsilon_{\alpha}<10^{2}$, and $<10^{-4}<\varepsilon_{i}<10^{2}$ 
in logarithmic steps.

\begin{figure}[t]
\begin{center}
\includegraphics[width=3in]{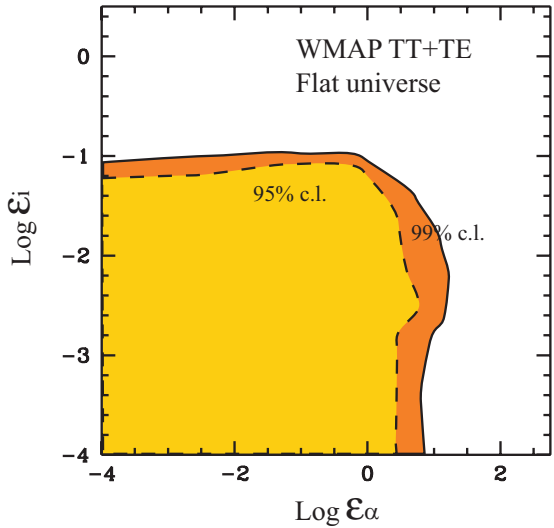}
\caption{Likelihood contour plot in the $\varepsilon_{\alpha}-\varepsilon_{i}$ 
plane showing the $2\sigma$ and $3\sigma$ contours.}
\label{fig4}
\end{center}
\end{figure}

The theoretical models are compared with the recent temperature and 
temperature-polarization WMAP data using the publicly available likelihood 
code \cite{WMAPanalysis03,kogut}.

In Fig. \ref{fig4} we plot the likelihood contours in the 
$\varepsilon_{\alpha}-\varepsilon_{i}$ plane showing the $2\sigma$ and 
$3\sigma$ contours. As we can see there is no strong correlation between 
the two parameters.

Marginalizing over all the remaining nuisance parameters we obtain 
the following constraints: $\varepsilon_{\alpha}< 10^{0.5}$ and 
$\varepsilon_{i}< 10^{-1.2}$ at $95\%$ c.l..

\begin{table}[t]
\renewcommand*{\arraystretch}{1.5}

\label{tab}
\begin{tabular}{|lll|c|}
\hline
&&&$\Delta\chi^{2}$ \\
\hline
\ \ PBH &$c_{a}=0.3$, &$c_{i}=0.13$&$1.9$
\\
 \ \ SHDM &$\Theta\le 10^{7}$,&$z<5.3$&$\le0.3$
\\& $\Theta= 10^{8},$& $z=6.0$ &$8.7$
\\
& $\Theta= 10^{9},$& $z=6.5 \ \ \ \ $ &$ \  3757 \ $\\
\hline
\end{tabular}
\caption{Difference in $\chi^{2}$ from the best fit standard recombination 
model for a range of non-standard recombination scenarios discussed in 
section \ref{sec2}, using WMAP TT and TE data. For the SHDM models we 
consider a range of decay lifetimes, $t_{x}$, measured in terms of the 
lifetime of the universe at a redshift, $z(t_{x})$.}
\end{table}
 \begin{figure}[t]
\begin{center}
\includegraphics[width=3in]{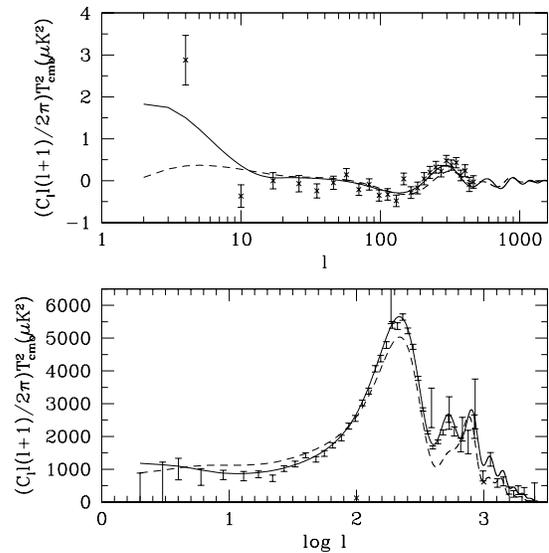}
\caption{Figures demonstrating how a potential degeneracy between flat and spatially curved scenarios with non-zero $\varepsilon_{\alpha}$ is broken with 
WMAP TT and TE spectra. A WMAP best fit fiducial model with $z_{ri}=17$ and $\Omega_{K}=0$ (full line) is shown 
against a model with
$\Omega_{K}=0.6$ and $\varepsilon_{\alpha}=10^{3}$ and $z_{ri}=17$ (dashed line) that prior to the WMAP data release had some degeneracy with the standard scenario. In particular note how the addition of the TE spectrum strongly breaks the degeneracy, since the open model has a strongly suppressed reionization peak.}
\label{fig5}
\end{center}
\end{figure}
 \begin{figure}[t]
\begin{center}
\includegraphics[width=3in]{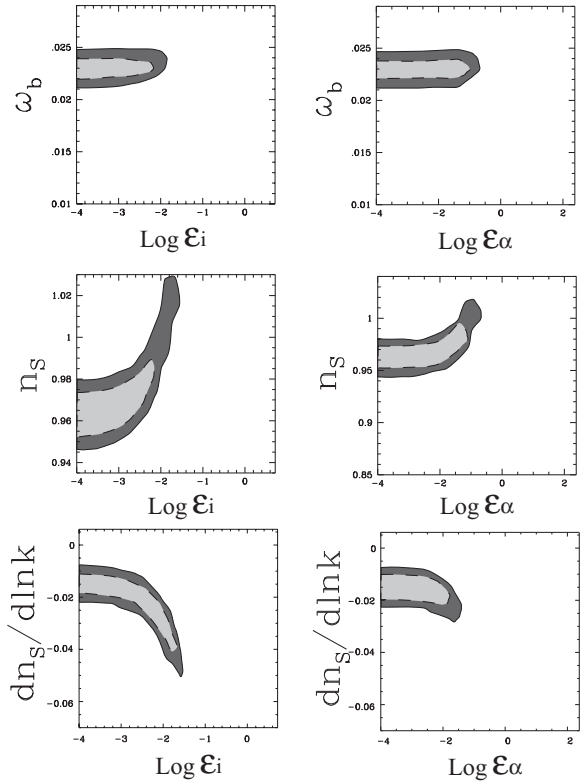}
\caption{Correlations between $\varepsilon_{\alpha}$ and $\varepsilon_{i}$ 
with several cosmological parameters. The redshift of reionization is fixed 
at $z_{ri}=12$ and the Hubble parameter is $h=0.68$. One can see that while the baryon density is robust to alterations in $\varepsilon_{\alpha}$ and $\varepsilon_{i}$, both the scalar spectral index and its running are sensitive.}\label{fig6}
\end{center}
\end{figure}

As we can see, despite the high precision of the WMAP data, substantial 
modifications to the recombination process might be present. 
In Table I we demonstrate this for models discussed in section 
\ref{sec2}.
Ionizing photons from PBH decay are still in good agreement 
with the data, to within $1\sigma$. Similarly, assuming the normalization 
prescription given in \cite{Dorosh02}, long lived SHDM particles are not 
constrained well by the data, however there is a sharp cut off in the 
likelihood for particles that would decay on timescales shorter than $t(z=6)$.

Since a modified recombination scheme can still be in agreement with the 
data, it is interesting to study the impact these modifications might have on the 
constraints of several parameters determined in the standard analysis.
In Fig. \ref{fig5} we demonstrate the benefits of having both the TT and TE 
spectra available in breaking potential degeneracies with spatial curvature. 
With pre-WMAP uncertainties in the TT spectrum 
peak heights relative to the plateau,  a `degeneracy' between curvature and 
$\varepsilon_{\alpha}$ existed \cite{Seager00}. A non-zero value of 
$\varepsilon_{\alpha}$  can shift the TT spectrum peak positions in an open 
model so that they have a similar a fiducial $\Lambda$ CDM case. 
However the accuracy of WMAP data over the first and second peaks, now 
allows allows such an open model to be distinguished by the peak suppression 
produced by increasing $\varepsilon$. Moreover, the  reionization peak 
is strongly suppressed in the open scenario so the two scenarios are further differentiated 
by the inclusion of TE data.

\begin{figure}[tt]
\begin{center}
\includegraphics[width=3in]{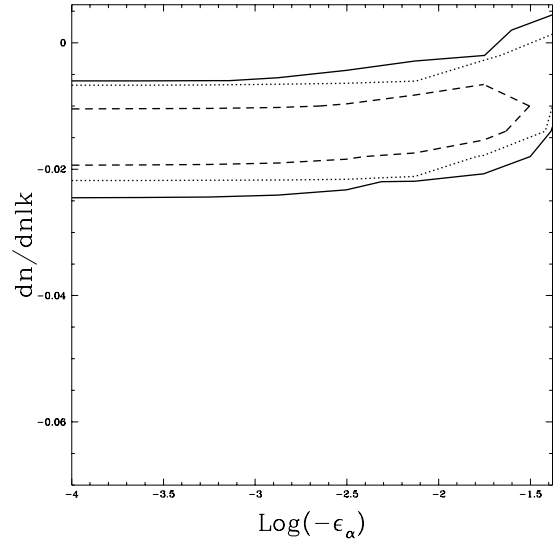}
\caption{Correlation between a negative $\varepsilon_{\alpha}$ 
and a running of the spectral index $dn/dlnk$ at $k=0.05 hMpc^{-1}$.
The redshift of reionization is fixed 
at $z_{ri}=12$, the Hubble parameter is $h=0.68$, $n_{s}=0.93$ and
$\varepsilon_{i} = 0$}
\label{fig7}
\end{center}
\end{figure}

In Fig. \ref{fig6} we probe degeneracies that are less easily distinguished; 
projecting the constraints on $\varepsilon_{\alpha}$ and $\varepsilon_{i}$ 
in function of the baryon density, the spectral index and its running 
$dn_{s}/dlnk$ at $k_0=0.05 h Mpc^{-1}$. 
As we can see, the constraints on $\omega_b$ are weakly affected by our 
simple modifications to the recombination process. 
The value of $\omega_b = 0.023\pm 0.001$ is unaffected by the inclusion of 
$\varepsilon_{\alpha}$ or $\varepsilon_{i}$. 
This can be explained by the high precision measurements of Sachs-Wolfe 
plateau and the first $2$ acoustic peaks made by the WMAP satellite, 
which breaks the degeneracy between $\omega_b$ and $\varepsilon_{i}$ 
found in Seager et al. Increasing $\varepsilon_{\alpha}$ or $\varepsilon_{i}$, 
however, can have an important impact on the determination of $n_s$. 
Namely, a modified recombination will have effects similar to those of an 
early reionization process, lowering the small scale anisotropies and allowing 
greater values of$n_s$ to be in agreement with the data.

Finally, it is interesting to study the correlation with a possible running of 
the spectral index. In Peiris et al. (\cite{peiris}), 
a $\sim 2 \sigma$ evidence for a 
negative running of the spectral index has been found. This is not expected 
in most common inflationary scenarios and several mechanisms have been proposed to explain the effect (see e.g. \cite{trodden}). However, as we can see from the plots at the bottom of Fig.\ref{fig6}, a delayed recombination will in general not solve this problem, but it will enhance it towards a more negative running 
(in the plots we assume $n_s=0.93$).  A possible solution might come from a speeding up of 
recombination rather than delay, through having concentrations of 
baryons in high density, low mass clouds, for example 
\cite{Naselsky02}. The accelerated recombination, because of the resultant premature reduction in the ionization fraction, can be parameterized by an effective $\epsilon_{\alpha}< 0$ 
with $\epsilon_{i}\ge 0$ still.
As we see in Fig.\ref{fig7}, negative values of $\epsilon_{\alpha}$
are more consistent with a zero running, although are likely not sufficient to fully explain the observed phenomenon.

\section{Conclusions}
\label{sec4}

We have probed the upper bounds that can be placed on the contribution of 
extra Ly-$\alpha$ and ionizing photon-producing sources through their effect 
on recombination and subsequently on the CMB  TT and TE spectra.  
We have found that, adopting a simple parametrization using constant,
effective values for $\varepsilon_{\alpha}$ and $\varepsilon_{i}$,
the WMAP data contraints $\varepsilon_{\alpha}<10^{0.5}$ and 
$\varepsilon_{i}<10^{-1.2}$ at the $95\%$ level.

We have found that the WMAP data is able to break many of the
worrying degeneracies between  $\varepsilon_{\alpha}$, $\varepsilon_{i}$
 and the more standard cosmological parameters. In particular,
the constraints on curvature and baryon density appear to be weakly affected
by a modified recombination scheme. However, it may affect the
current WMAP constraints on inflationary parameters like the spectral
index $n_{s}$ and its running.

Physically motivated models, like those based on primordial
black hole or super heavy dark matter decay, are able
to provide a good fit to the current data. 
Future observations in both temperature and polarization, from the next WMAP release and the Planck satellite \cite{Dorosh02}, will be needed if we are to more stringently test these models.

\medskip 
\textbf{Acknowledgements} It is a pleasure to thank David Spergel 
for helpful comments and suggestions. RB is supported by WMAP.  
We acknowledge the use of CMBFAST~\cite{CMBFAST}. 
We thank the organizers of the CMBNET workshop in Oxford, February 2003, 
and the Oxford-Princeton Workshop on Cosmology, in Princeton, March 2003, 
where portions of this work were completed.

\end{document}